\documentclass[preprint,superscriptaddress,amsmath,amssymb,aps,prl,]{revtex4-1}

\usepackage{color}
\usepackage{graphicx}
\usepackage{dcolumn}
\usepackage{bm}

\begin{document}


\title{Giant Anomalous Hall Effect due to Double-Degenerate Quasi Flat Bands}

\author{Wei Jiang}
\email{jiangw@umn.edu}
\affiliation{Department of Electrical $\&$ Computer Engineering, University of Minnesota, Minneapolis, Minnesota 55455, USA.}

\author{Duarte J. P. de Sousa}
\affiliation{Department of Electrical $\&$ Computer Engineering, University of Minnesota, Minneapolis, Minnesota 55455, USA.}

\author{Jian-Ping Wang}
\affiliation{Department of Electrical $\&$ Computer Engineering, University of Minnesota, Minneapolis, Minnesota 55455, USA.}

\author{Tony Low}
\email{tlow@umn.edu}
\affiliation{Department of Electrical $\&$ Computer Engineering, University of Minnesota, Minneapolis, Minnesota 55455, USA.}

\date{\today}

\begin{abstract}

We propose a novel approach to achieve giant AHE in materials with flat bands (FBs). FBs are accompanied by small electronic bandwidths, which consequently increases the momentum separation ($K$) within pair of Weyl points and thus the integrated Berry curvature. Starting from a simple model with a single pair of Weyl nodes, we demonstrated the increase of $K$ and AHE by decreasing bandwidth. It is further expanded to a realistic pyrochlore lattice model with characteristic double degenerated FBs, where we discovered a giant AHE while maximizing the $K$ with nearly vanishing band dispersion of FBs. We identify that such model system can be realized in both pyrochlore and spinel compounds based on first-principles calculations, validating our theoretical model and providing a feasible platform for experimental exploration. 

\end{abstract}

\maketitle

\paragraph*{Introduction} The anomalous Hall effect (AHE), i.e., a zero field Hall conductivity observed in ferromagnetic materials mediated by spin-orbit coupling, is one of the most intriguing electronic transport phenomena~\cite{Nagaosa2010}. It has been proposed for magnetic sensors and memories for their high sensitivity and thermal stability~\cite{Moritz2008,Lu2012,Wang2020} and energy efficient spintronics applications~\cite{Wang2017,Mahendra2018,zhang2019,Puebla2020}. Two prevalent theories explain the experimentally observed AHE, i.e., intrinsic AHE due to spin-orbit coupling and extrinsic AHE due to impurity scattering, such as side jump and skew scattering mechanisms~\cite{Smit1955,Berger1970,Nagaosa2010,Ye1999,Yao2004,Onoda2006}. The theory of intrinsic AHE was first put forth by Karplus and Luttinger~\cite{Karplus1954}, and was subsequently better appreciated due to Berry curvature of the occupied Bloch bands ~\cite{Ye1999,Jungwirth2002,Yao2004,Xiao2010}. More recently, with the discovery of various topological states, Weyl semimetal systems have been touted as fertile ground for large AHE as Weyl points and their vicinity can intrinsically host large Berry curvature~\cite{Burkov2011,Burkov2014,Zyuzin2016,Jiang2020}. Large AHE has indeed been observed in several material candidates that have been studied both theoretically and experimentally~\cite{liu2018,Wang2018,Shekhar2018}. Several rule of thumb have been suggested for achieving high AHE, however, there is no consensus on their general applicability~\cite{Wang2018,Shekhar2018,Derunova2019}.

Theory has elaborated that the intrinsic peak anomalous Hall conductivity (AHC) in Weyl semimetals with single pair of Weyl nodes is given by $\sigma_{xy}=\frac{e^2K}{4\pi^2}$, where $K$ is the momentum separation between the Weyl nodes~\cite{Burkov2014}. To maximize AHE, an apparent approach would be to increase $K$, which in principle, could be tuned through band engineering. In general material systems, there are usually multiple pairs of Weyl nodes, which are mostly located at different energy due to the dispersing bands. As a consequence, their corresponding AHC as a function of energies usually shows multiple peaks and the Fermi level is rarely coincident with the AHC peak chemical potential. In this work, we show that the simultaneous harvesting of maximal Berry curvatures from multiple pairs of Weyl points can be achieved through the engineering of flat bands (FB)~\cite{Mielke1991,Liu2014,Jiang2019}. These FBs present limited electronic bandwidth where the Fermi level and peak AHC coincides. In addition, FBs also maximize the Weyl points separation $K$. 

Lowly dispersive bands could be achieved by tuning the inter-atomic hopping through strain engineering. However, the maximum allowable strain is limited by the materials' mechanical properties. FB systems would be ideal considering their intrinsic dispersionless bands. The pinning of the Fermi level to the FB would also be easier to achieve due to its large density of states. Strongly-correlated FB in low dimensions (1D or 2D) has been extensively studied, yielding various exotic quantum states, e.g., ferromagnetism, superconductivity, and topological states~\cite{Mielke1991,Liu2014,Jiang2019,Wu2007,Tang2011,Budich2013,Maimaiti2017,Jiang2019a}. Recently, a few three-dimensional (3D) FB systems have been proposed~\cite{Nishino2005,Guo2009,Weeks2012}, based on which various topological states are identified and demonstrated in pyrochlore and spinel compounds~\cite{Guo2009,Weeks2012,Hase2018,Zhou2019,Azadani2020}. However, to the best of our knowledge, there are no study of AHE in these systems, let alone the exploitation of these systems for optimal AHE. 

Here, we first elaborate a general mechanism to increase $K$ and thus AHC based on a model study with a single pair of Weyl nodes. Thereafter, we propose that FB in 3D systems can be a fertile ground for giant AHE where quasi double degenerate FBs can host the formation of multiple pairs of Weyl nodes at nearly the Fermi energy with optimized $K$. To elucidate on this idea, we begin with the fundamental pyrochlore lattice model that hosts 3D FBs, and study the influence of various hopping parameters to the FB based on the tight-binding model. Subsequently, intrinsic AHE is studied, where the characteristic features of FB on AHE are discussed and analyzed in terms of their Berry curvatures. Finally, we examine the validity of this model in both pyrochlore and spinel compounds hosting 3D FBs through DFT calculations, and demonstrate sizeable AHC around the FBs. Our work outlines a novel approach in the search of giant AHE materials for energy efficient spintronics applications~\cite{Mahendra2018,zhang2019,Puebla2020}.

\begin{figure}
\includegraphics[width=0.8\linewidth]{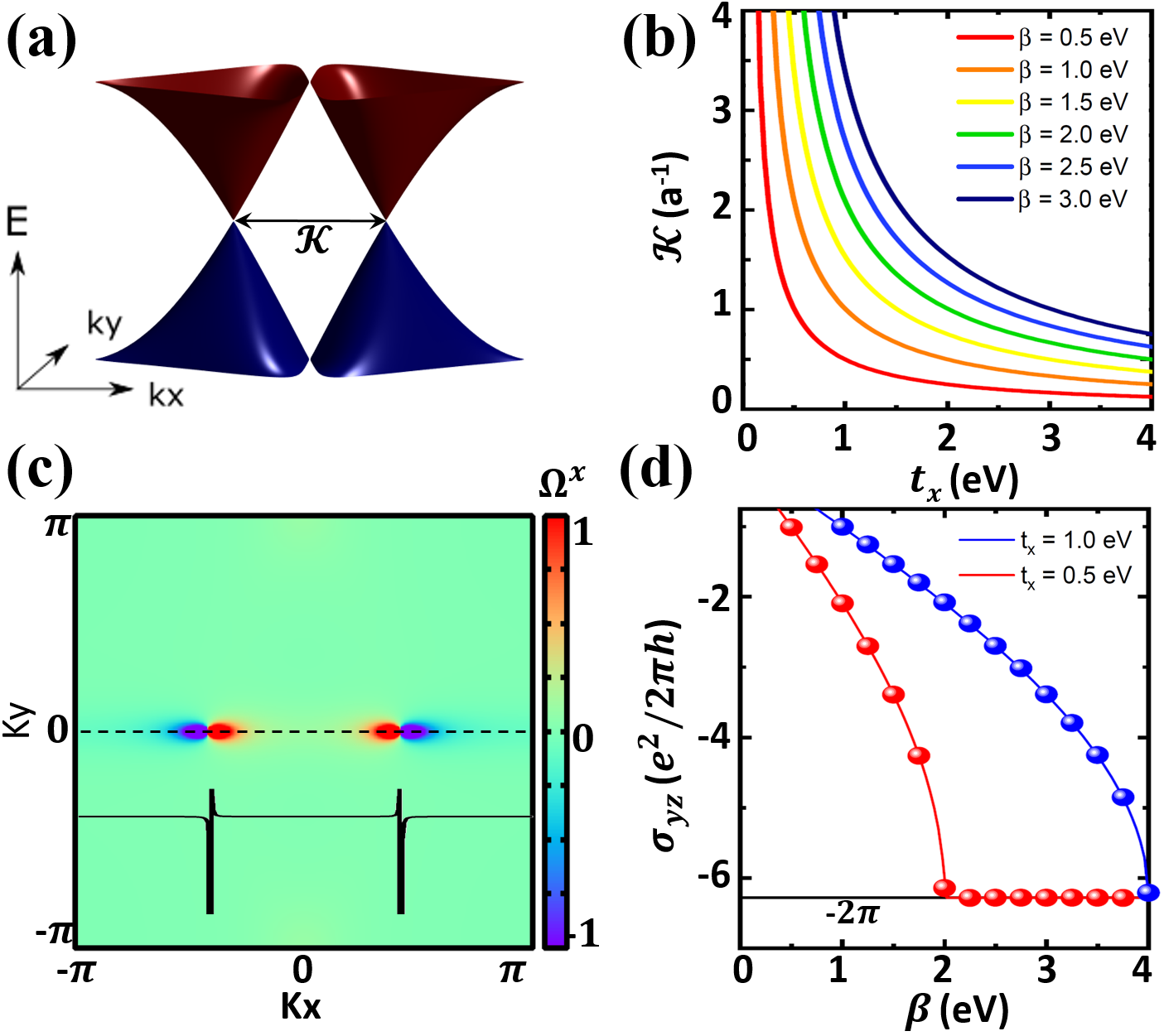} 
\caption{\textbf{Band structure and anomalous Hall conductivity of generic Weyl semimetals.} (a) Typical band structure of a magnetic Weyl semimetal derived from Eq.~(\ref{eq:Hm_toy}). We set an exchange field along the $x$ direction and the following parameters: $\beta = 2$ eV, $t_x = t_y = t_z = 1$ eV and $a = 1$ \AA\ . (b) Weyl node separation as a function of $t_x$ for several $\beta$ values. (c) Berry curvature ($\Omega^x$) distribution at the $k_x-k_y$ plane with inset showing linear scan along the dashed line. (d) Calculated AHC based on  Eq.~(\ref{eq:sigma_yz}) (symbols) and Weyl nodes separation ($\sigma_{xy}=\frac{e^2K}{4\pi^2}$, solid curves) as a function of the exchange splitting strength $\beta$ for different $t_x$ values. The horizontal line highlights the maximum value $-2\pi$ for which $\sigma_{yz}= e^2 /h$.}
\label{fig:fig0}
\end{figure}

\paragraph*{Weyl semimetal toy model} We start from the cleanest scenario with one single pair of Weyl nodes using a two-orbitals spin dependent cubic lattice model with the Hamiltonian given by~\cite{Vazifeh2013,Istas2019}    
\begin{eqnarray}
& \mathcal{H} = \tau_z \otimes [\textbf{f}(\textbf{k})\cdot \boldmath{\sigma}] + \tau_x \otimes [g(\textbf{k}) \sigma_0] + \tau_0 \otimes [(\beta/2)\sigma_x],
\label{eq:Hm_toy}
\end{eqnarray}
where $\textbf{f}(\textbf{k}) = \hat{x}t_x\sin(k_x a)+ \hat{y} t_y\sin(k_y a)+ \hat{z}t_z\sin(k_z a)$ and $g(\textbf{k}) = t_x (1 - \cos(k_x a)) + t_y (1 - \cos(k_y a)) + t_z (1 - \cos(k_z a))$ are the structure factors;  $\boldmath{\sigma} = \hat{x} \sigma_x + \hat{y} \sigma_y + \hat{z} \sigma_z$ is the vector of Pauli matrices; $a$ is the lattice constant and $t_{i}$ is the nearest neighbor hopping parameter along the $i$-th axis with $i=x,y,z$. The Pauli matrices $\tau$/$\sigma$ operate in the orbital/spin space and $\beta$ defines the exchange splitting strength. Figure~\ref{fig:fig0}(a) displays a typical band structure derived from Eq.~(\ref{eq:Hm_toy}) for $\beta = 2$ eV and $t_x = 1$ eV, which shows a pair of Weyl nodes located at $\boldmath{k} = (\pm k_0, 0, 0)$ with $k_0 = \arccos(1-\beta^2/(8 t_x^2))$ (see details in Supporting Information \cite{Supp}). In particular, the momentum separation between the two Weyl nodes, $K = 2k_0$, increases with the increase of the exchange splitting $\beta$ and/or the decrease of the hopping parameter $t_x$ along the magnetization direction. Such behavior is evident in Fig.~\ref{fig:fig0}(b) where $K$ is plotted as a function of $t_x$ for several $\beta$ values.

It was previously shown that the AHC of a Weyl semimetal increases linearly with $K$~\cite{Burkov2014}. The intrinsic AHC can be calculated by simply integrating Berry curvature of the occupied Block states, as prescribed by the Kubo formula~\cite{Yao2004}: 
\begin{eqnarray}
\Omega_n^x(\textbf{k})= -\sum_{n^{'}\neq n} \frac{2\mathrm{Im}[\langle\phi_{n\textbf{k}}\mid v_y\mid\phi_{n'\textbf{k}}\rangle\langle\phi_{n'\textbf{k}}\mid v_z\mid\phi_{n\textbf{k}}\rangle ]}{(\epsilon_{n'\textbf{k}}-\epsilon_{n\textbf{k}})^2},
\label{eq:Berry_x}
\end{eqnarray}
\begin{eqnarray}
\sigma_{yz}=-\frac{e^2}{\hbar}\int_{BZ}\frac{d^3k}{(2\pi)^3}\sum_{n}f_{n\textbf{k}}\Omega_n^x(\textbf{k}),
\label{eq:sigma_yz}
\end{eqnarray}
where $f_{n\textbf{k}} = \Theta(\epsilon_F - \epsilon_{n\textbf{k}})$ is the zero temperature Fermi-Dirac distribution, $v_{y,z}$ refers to the velocity operator, $\epsilon_{n\textbf{k}}$ and $\mid\phi_{n\textbf{k}}\rangle$ are the $n$-th energy band and the associated Bloch eigenstate, respectively. To better understand how Berry curvature of Weyl pair contributes to the AHC, we show the Berry curvature distribution, $\Omega^x$, at the $k_x-k_y$ plane when the Fermi energy $\epsilon_F$ coincides with the Weyl nodes in Fig.~\ref{fig:fig0}(c). The Berry curvature is mainly distributed in the vicinity of the Weyl nodes with alternating positive and negative values at opposite sides of the Weyl nodes along the Weyl pair separation direction. The non-compensating Berry curvature near each Weyl nodes contributes to the AHC, as can be seen from the linear scan of Berry curvature along the dashed line across the Weyl pair [inset of Fig..~\ref{fig:fig0}(c)\cite{Supp}]. To verify the relation, $\sigma_{xy}=\frac{e^2K}{4\pi^2}$, we explicitly calculate AHC based on Eq.~(\ref{eq:sigma_yz}) as a function of $\beta$ for different $t_x$ values, as presented with symbols in Fig.~\ref{fig:fig0}(d). This agrees perfectly with the AHC calculated based on $K$ using $\sigma_{xy}=\frac{e^2K}{4\pi^2}$, as displayed with solid lines in Fig.~\ref{fig:fig0}(d). As anticipated, the AHC becomes larger with increasing $\beta$ and could even reach the quantum limit of $e^2 /h$ when $K = 2\pi /a$. The results suggests that smaller $t_x$ could also enhance AHC, as anticipated since smaller $t_x$ implies larger $K$.

\begin{figure}
\includegraphics[width=\linewidth]{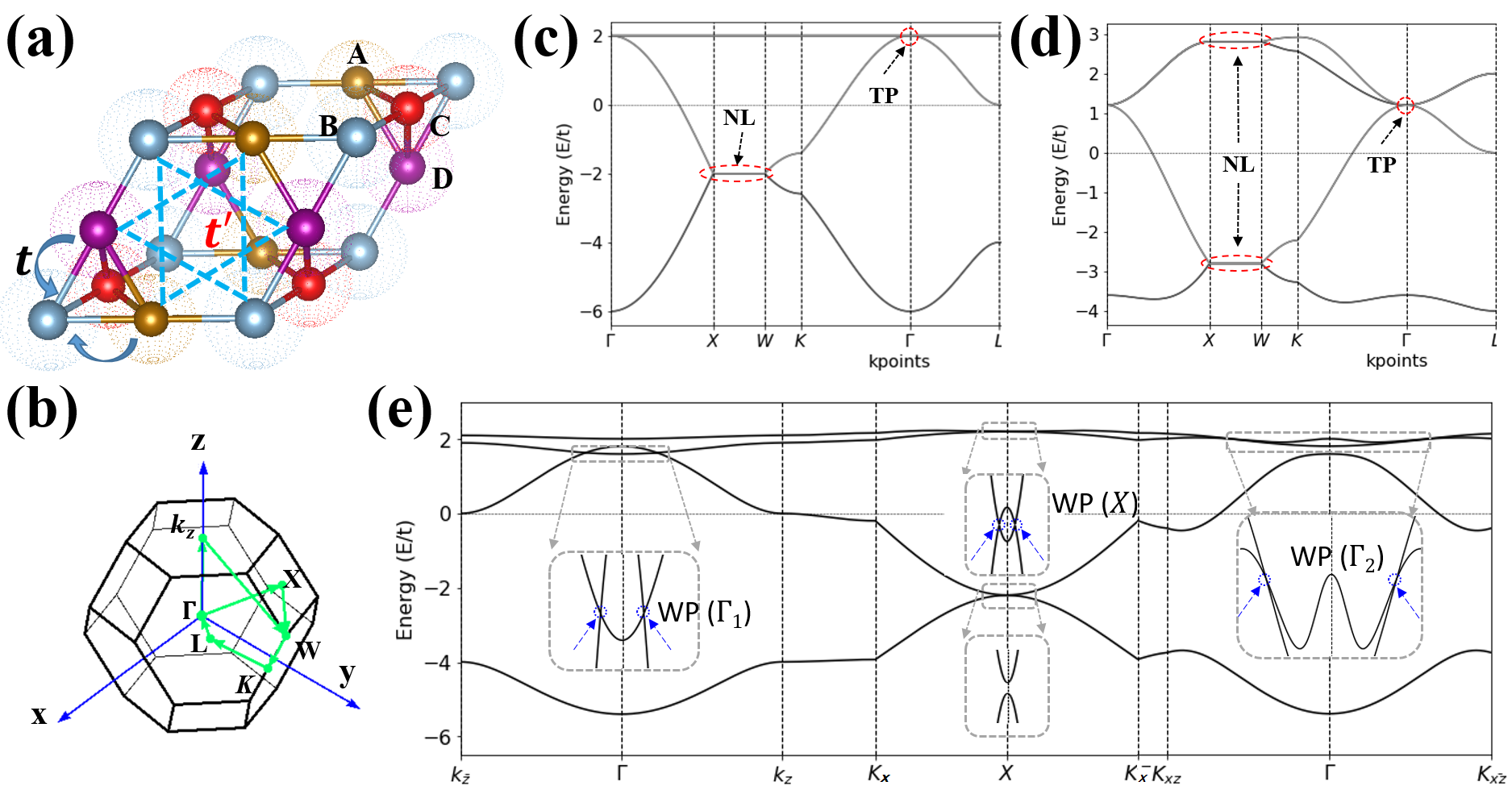} 
\caption{\textbf{Band structure of pyrochlore lattice.} (a) Crystal structure of pyrochlore lattice with four atoms (A-D) in one unit cell. $t$ and $t'$ indicate the NN and NNN hoppings, respectively. (b) High symmetry k-path in the first Brillouin zone. (c) Band structure of ideal pyrochlore lattice without considering NNN interaction and SOC effect. (d) Band structure with NNN interaction $t'$ = 0.2$t$. The triple degenerate point (TP) and nodal lines (NL) are highlighted by red ellipses. (e) Band structure considering SOC ($\lambda$ = -0.2$t$) and broken TRS ($\lambda_z$ = 5$t$) with different Weyl points (WPs) highlighted.}
\label{fig:fig1}
\end{figure}

\paragraph*{Pyrochlore lattice tight-binding model} Beyond the toy model, we extend our study to more realistic material systems, where we can further verify the aforementioned mechanism and also explore promising material candidates. From band structure perspective, smaller $t_x$ usually leads to narrower bands. The extreme scenario would be the FB case. Normally, the formation of FB in 3D is related to trivial defect states, such as point defect or dislocations~\cite{Corsetti2011,Hu2018}. Conversely, the 3D FB of interest in this work requires destructive interference that leads to zero net hopping despite nonvanishing hopping terms throughout the lattice. This is very stringent for a real lattice model and there are limited studies of 3D FB ~\cite{Guo2009,Weeks2012,Hase2018} compared to their 1D or 2D counterparts~\cite{Mielke1991,Liu2014,Jiang2019,Wu2007,Tang2011,Budich2013,Maimaiti2017,Jiang2019a}. Here, we will focus on the pyrochlore lattice, named after pyrochlore compounds~\cite{Chakoumakos1984,Gardner2010}, which is a 3D network of corner-sharing tetrahedron, as shown in Fig.~\ref{fig:fig1}(a). Pyrochlore lattice, with the group symmetry of $Fd\Bar{3}m$, can also be viewed as stacking of tetrahedron clusters in the hcp lattice, where each tetrahedron contains four atoms in each unit cell (A, B, C, and D in Fig.~\ref{fig:fig1}). Each atom has six nearest neighbors (NNs) and twelve next nearest neighbors (NNNs), as indicated by $t$ and $t'$ in Fig.~\ref{fig:fig1}(a), respectively. Considering one orbital on each atomic site, the system can be described by a four- or eight-band Hamiltonian depending on whether spin degree of freedom is considered. We limit our Hamiltonian to only the essential NN and NNN hoppings, which are sufficient to demonstrate the related physics and can be written as:
\begin{equation}
\mathcal{H} = \sum_{i\sigma} \epsilon_i d_{i\sigma}^\dagger d_{i\sigma} -t \sum_{\langle i,j\rangle \sigma} d_{i\sigma}^\dagger d_{j\sigma} - t'\sum_{\langle \langle i,j\rangle\rangle\sigma}  d_{i\sigma}^\dagger d_{j\sigma} + H.c. ,
\label{eq:Lieb}
\end{equation}
where $\epsilon_i$ represents the on-site energy at site $i$; $d_{i\sigma}^\dagger$ and $d_{i\sigma}$ are the creation and annihilation operators of electrons at site $i$ with spin $\sigma$, respectively; $t$ and $t'$ are the NN and NNN hoppings, respectively. 

We first analyze the ideal case with zero NNN hopping $t'$ and uniform on-site energy $\epsilon_i=0$ and NN hopping $t$. Given the spin degeneracy, one copy of the momentum space Hamiltonian can be obtained through $\mathcal{H}=\sum_\textbf{k} \Psi_\textbf{k}^\dagger H(\textbf{k})\Psi_\textbf{k}$, where $\Psi_\textbf{k}^\dagger=(d_{A\textbf{k}}^\dagger,d_{B\textbf{k}}^\dagger,d_{C\textbf{k}}^\dagger,d_{D\textbf{k}}^\dagger)$. The diagonalization of Hamiltonian $H(\textbf{k})$ yields four eigenstates with two degenerate FBs $E^{1,2}=2t$ and two dispersive bands $E^{3,4}=-2t(1\pm\sqrt{1+A_k})$, where $A_k=cos(2k_x)cos(2k_y)+cos(2k_x)cos(2k_z)+cos(2k_y)cos(2k_z)$. One of the dispersive bands touches the FBs at the $\Gamma$ point forming a triple degenerate point (TP), as shown in Fig.~\ref{fig:fig1}(c) for positive $t$ scenario. Two dispersive bands form nodal lines (NL) along X-W and its symmetry invariant k-paths, i.e., diagonals of the square faces of the Brillouin zone [Fig.~\ref{fig:fig1}(b) and (c)]. We further consider the NNN hopping effect \cite{Supp}, the two FBs become dispersive and disperse upwards/downwards for positive/negative $t'$, as shown in Fig.~\ref{fig:fig1}(c) for the positive $t'$ case. The evolution of the band structure with different NNN hopping strength $t'$ is also plotted \cite{Supp}, showing the increase of bandwidth of FBs with increasing $|t'|$. The degeneracy of the two FBs is lifted at k-paths with low symmetry, however, the quadratic band touching at the $\Gamma$ point with TP and the NL feature remain robust [Fig.~\ref{fig:fig1}(d)]. 
To induce the magnetic Weyl semimetal phase, we further add a NNN SOC term ($H_{SOC}$) and a Zeeman type exchange splitting term along the $z$ direction ($H_z$) to our Hamiltonian:
\begin{equation}
    H_{SOC}=i\lambda\sum_{\langle \langle i,j\rangle\rangle\alpha\beta}(\overrightarrow{r_{ij}^1}\times\overrightarrow{r_{ij}^2})\cdot\sigma_{\alpha\beta}s_{i\alpha}^\dagger s_{j\beta},
\end{equation}
\begin{equation}
    H_z = \lambda_z\sum_{i\alpha}d_{i\alpha}^\dagger\sigma_z d_{j\alpha},
\end{equation}
where $\lambda$ and $\lambda_z$ are used to describe the SOC coupling and the exchange coupling strength, respectively. $\overrightarrow{r_{ij}^{1,2}}$ are the NN vectors that traverse between NNN sites $i$ and $j$, and $\sigma$ is the Pauli spin matrices. When $\lambda_z = 0$, time reversal symmetry (TRS) is preserved, the band structure shows that the TP at $\Gamma$ point is splitted into one double degenerate and one single state, and the NL degeneracy is also lifted leaving only X point degenerate \cite{Supp}. Various topological states evolves due to the SOC effect, i.e., topological Dirac semimetal at one/three fourth filling and topological insulator at one half filling for $\lambda=-0.2t$; topological semimetal states for $\lambda=0.2t$ \cite{Supp}. After breaking the TRS with nonzero $\lambda_z$, the system experiences a series of topological phase transitions with different $\lambda_z$~\cite{Zhou2019}, e.g., Chern insulator and magnetic Weyl semimetal states at large $\lambda_z$ limit, as shown in Fig.~\ref{fig:fig1}(e). It is more intuitive to understand the formation of the Weyl points (WPs) by considering the evolution of the Dirac points after breaking TRS, e.g., the WPs($\Gamma_1$) along $k_z-\Gamma$ path arises from the DP at $\Gamma$ point and similar for WP(X) and WP($\Gamma_2$) that arise from DP at X and $\Gamma$, respectively, as shown in Fig.~\ref{fig:fig1}(e) \cite{Supp}. It is worth mentioning that the NLs are completely gapped due to the broken TRS, corresponding to a Chern insulator.

\begin{figure}
\includegraphics[width=0.8\linewidth]{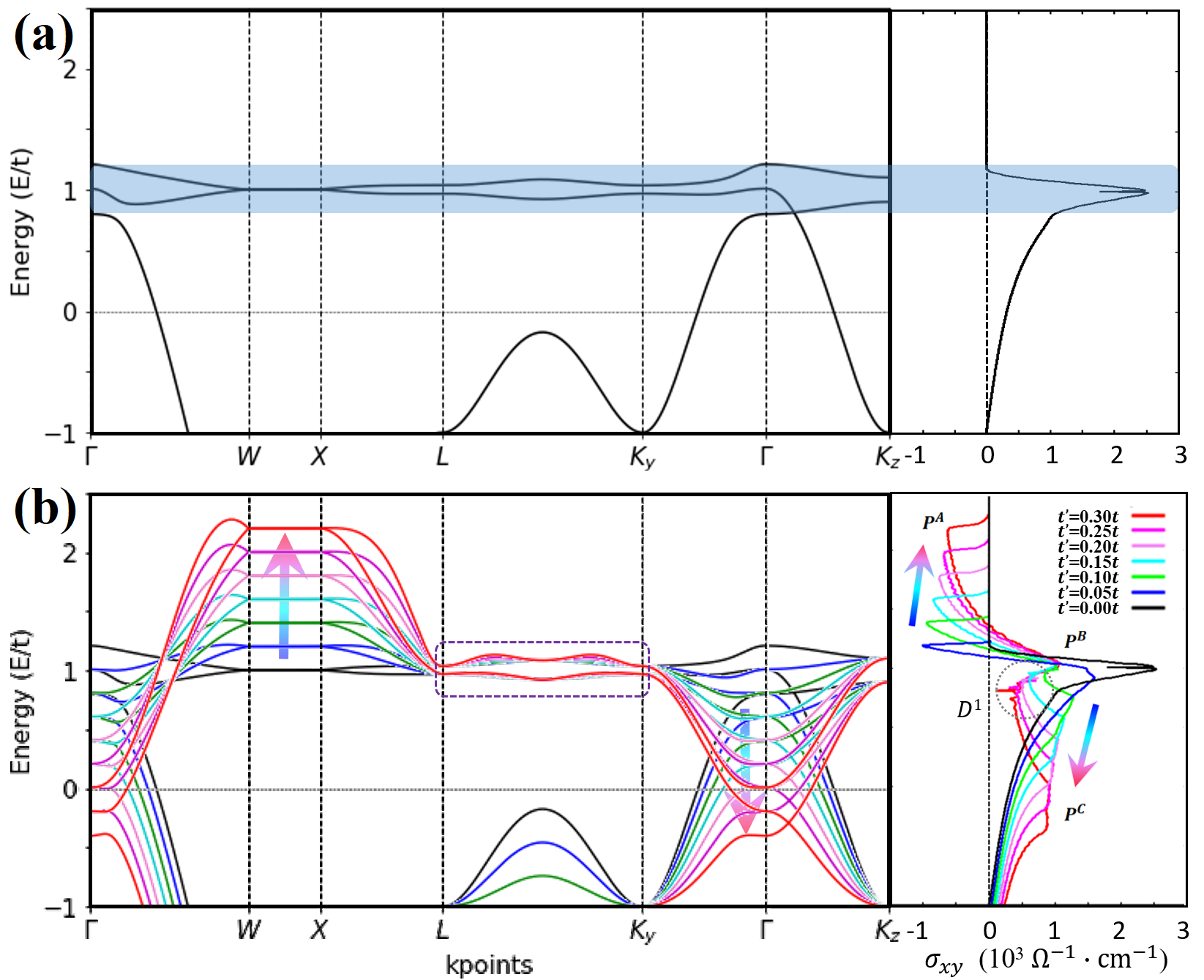} 
\caption{\textbf{Evolution of AHC with the change of NNN hopping strength $t'$.} (a) Left and right panel shows band structure and AHC of the pyrochlore model with zero $t'$. (b) Same as (a) for evolution of the band structure and AHC with the change of NNN hopping strength $t'$, respectively. AHC decreases with the increase of the band width due to the increase with the $t'$, as indicated by the colored arrow.}
\label{fig:fig2}
\end{figure}

\paragraph*{Anomalous Hall Effect} With broken TRS in conjunction with SOC, various Weyl points are created near the FBs that serve as hotspots for Berry curvature. Since the intrinsic AHC is purely the sum of Berry curvatures, we can calculate the AHC ($\sigma_{xy}$) based on Eq.~(\ref{eq:sigma_yz}). For simplicity, we will only consider the large exchange splitting limit, i.e. $\lambda_z=5t$. The upper branch of the band structure and AHC for the case when $t'=0,\lambda=0.2t$ is shown in Fig.~\ref{fig:fig2}(a). The lower branch of the band structure (not shown) is symmetric to that of the upper branch, with the same AHC but opposite sign \cite{Supp}. Clearly, there is a large peak right at the energy of the FBs, where multiple pairs of Weyl nodes are residing. To understand the correlation between flatness of FBs and the giant AHC, we studied the evolution of AHC with different band dispersion by tuning the NNN hopping term $t'$ in our TB model. The evolution of band structure and AHC with the change of $t'$ is shown in Fig.~\ref{fig:fig2}(b). Evidently, with the increase of $t'$, the bandwidth of the `FBs' increases with a pronounced change in AHC. The giant single AHC peak splits into three different peaks with a noticeable degradation in the maximum AHC, as indicated by the arrows in Fig.~\ref{fig:fig2}(b). These results are consistent with our conjecture that the giant AHC is closely related to the flatness of the `FB' as elucidated by the toy model. 

By analyzing the evolution of band structure and AHC [Fig.~\ref{fig:fig2}(b)], we notice that the energy of one of the splitted peaks, $P^A$, is shifting to higher energies with increasing $t'$. Its energy is always coincident with that of high-symmetry k-path $X-W$, which happens to be highly energetically degenerate with almost zero dispersion. Another peak, $P^B$, remains at the same energy level that corresponds to the energy of $L-K_y$ and its symmetry invariant k paths. From the band structure highlighted by dashed square in Fig.~\ref{fig:fig2}(b), we find the two bands remain nearly flat with very small energy difference along these k-paths \cite{Supp}. Similarly, we find that the position of peak $P^C$ is coincident with the position of the crossing point between bottom FB and upper Dirac band along $K_z$ and $\Gamma$ k-path. Different from $P^A$, $P^C$ is much broader and the position of $P_C$ shifts downward with the increasing $t'$. We note that there is also a small dip ($D^1$) between $P^B$ and $P^C$, which also shifts to the lower energy with increasing $t'$.

\begin{figure}
\includegraphics[width=\linewidth]{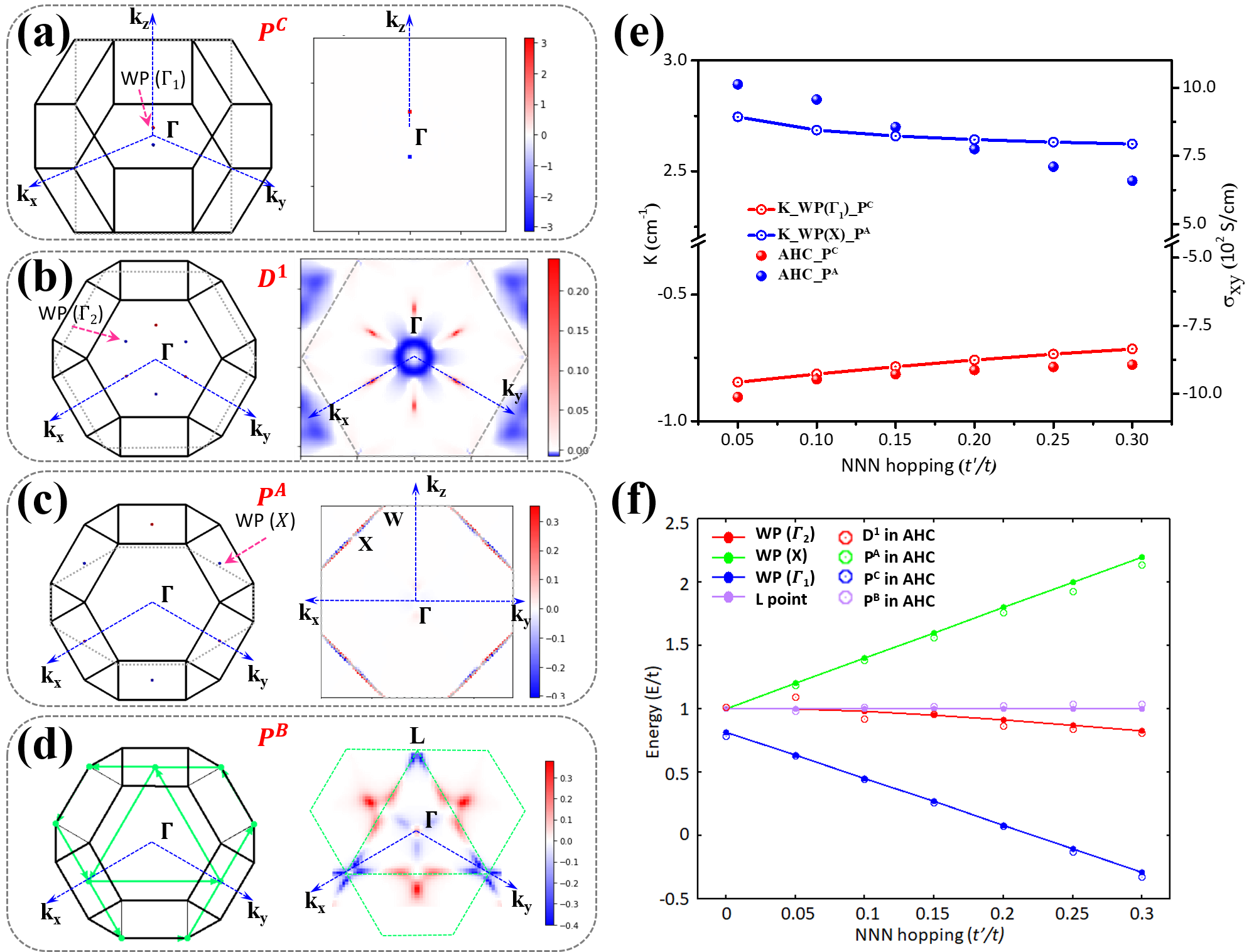} 
\caption{\textbf{Distribution and evolution of the Weyl nodes and Berry phase with different $t'$.} (a) Distribution of the Berry phase in the $xz$ plane for $P^C$ (right panel), showing the contribution from the pair of Weyl points (left panel), WP($\Gamma_1$), with positive and negative chirality. (b) Same as (a) for WP($\Gamma_2$). Right panel shows the distribution of Berry phase in the (001) plane across the $\Gamma$ point. (c) Same as (a) for the  P$^A$ due to WP(X) and Berry phase distribution in the (110) plane across the $\Gamma$ point. (d) Same as (a) for the P$^B$ and the Berry phase distribution in the (001) plane through $L-K_x-K_y$. (e) Evolution of $K$ for WP($\Gamma_1$) and WP(X) and their corresponding AHC, $P^{C,A}$ with different $t'$. (f) Energy evolution of WP($\Gamma_1$, X, $\Gamma_2$) and L point in comparison with evolution of peaks $P^{A,B,C}$ and $D^1$ with different $t'$.}
\label{fig:fig3}
\end{figure}

To better understand the underlying mechanism and further validate the connection between the band structure and AHC, we calculate the position of Weyl nodes and Berry curvature related to $P^{A,B,C}$ and $D^1$. The band crossing points between upper Dirac and lower FB are well isolated from the other bands in the energy space, which corresponds to an ideal magnetic Weyl semimetal with one single pair of Weyl nodes~\cite{Zhou2019}. The Weyl nodes are distributed along the $K_z$ direction symmetrically with respect to the $\Gamma$ point, as shown in Fig.~\ref{fig:fig3}(a) and WP($\Gamma_1$) in Fig.~\ref{fig:fig1}(e). As expected, the Berry curvature is mainly distributed around these two Weyl nodes that contribute to $P^C$, as shown in Fig.~\ref{fig:fig3}(a). We note that only $z$ component of the Berry curvature is presented due to the magnetization direction along $z$ direction. The broadening of the $P^C$ is due to the contribution of Berry curvature near the $\Gamma$ points \cite{Supp}. 

There exist another six pairs of Weyl nodes formed by the two `FBs', among which three pairs are located symmetrically around the $k_z$ axis along six directions near the (001) plane [Fig.~\ref{fig:fig3}(b) and WP($\Gamma_2$) in Fig.~\ref{fig:fig1}(e)]. The Berry curvature of the (001) plane crossing the $\Gamma$ point shows large contributions from these Weyl points, as shown in Fig.~\ref{fig:fig3}(b). Possibly due to the type-II feature of these Weyl fermions, [WP($\Gamma_2$) in Fig.~\ref{fig:fig1}(e)], it contributes negatively to the AHC (D$^1$)~\cite{Zyuzin2016}. The other three pairs of Weyl nodes are near $X$ and its symmetry-invariant $k$ points [Fig.~\ref{fig:fig3}(c) and WP(X) in Fig.~\ref{fig:fig1}(e)].  Similarly, significant Berry curvature contributions to $P^A$ can be seen along $X-W$ paths from Fig.~\ref{fig:fig3}(c), consistent with the very close energy between two FBs along those k-paths. Interestingly, corresponding to the $P^B$, there are also large Berry curvature contributions from $L$ and the middle of $L-K_y$ path even without existence of Weyl nodes, as shown in Fig.~\ref{fig:fig3}(d). The almost constant energy of $L$ point with different $t'$ is consistent with the evolution of $P^B$.

To verify the linear relationship between momentum separation of Weyl pairs and AHC, $\sigma_{xy}=\frac{e^2K}{4\pi^2}$, we extract the momenta separation within pair of Weyl points (line with circles) in comparison with its AHC (solid circles) with changing $t'$, as shown in Fig.~\ref{fig:fig3}(e) \cite{Supp}. The evolution of WP($\Gamma_1$)/WP(X) show very good agreement with the change of AHC, $P^C$/$P^A$. The opposite sign in AHC for $P^C$ and $P^A$ is due to the opposite distribution of positive and negative Weyl nodes along the $k_z$ direction. We note that because of the type-II feature of WP($\Gamma_2$) and the close distribution between $D^1$ and $P^B$, such relationship is not applicable to WP($\Gamma_2$). We also plot the energy evolution of those Weyl nodes (line with dots) and AHC peaks (open circles), as shown in Fig.~\ref{fig:fig3}(f). The perfect agreement further verifies the contribution from each type of Weyl nodes to those AHC peaks. More importantly, when $t'$ becomes smaller, energies of Weyl nodes for $P^A$, $P^B$ and $D^1$ are closer to each other and that of $L$ point, leading to the giant AHC observed for $t'=0$. This suggest the superimposing of multiple pairs of Weyl nodes due to the FBs with small band dispersion can indeed enhance the AHC. Developing an understanding to the control of the relative signs of AHC from the multiple pairs of Weyl points would further enhance this giant AHC.

With this tunable pyrochlore lattice model, we can also theoretically study the SOC and structural effect on the AHE with different SOC ($\lambda$) strength and lattice constant ($a$). It is generally believed that large SOC is required to achieve large AHC~\cite{Jungwirth2002,Yao2004}. Due to that reason, searching of large AHE materials is usually limited to heavy metals. To study the SOC effect in our model, we also calculate the change of AHC with different SOC strength, $\lambda$. Surprisingly, the results show that the SOC has limited influence to the AHC, which is mainly contributed by the Weyl nodes \cite{Supp}. In addition, we also studied the effect of lattice constant, which shows smaller lattice will yield larger AHC \cite{Supp}. This is consistent with $\sigma_{xy}=\frac{e^2K}{4\pi^2}$, where $K$ depends inversely with the real space lattice constant.

\begin{figure}
\includegraphics[width=0.9\linewidth]{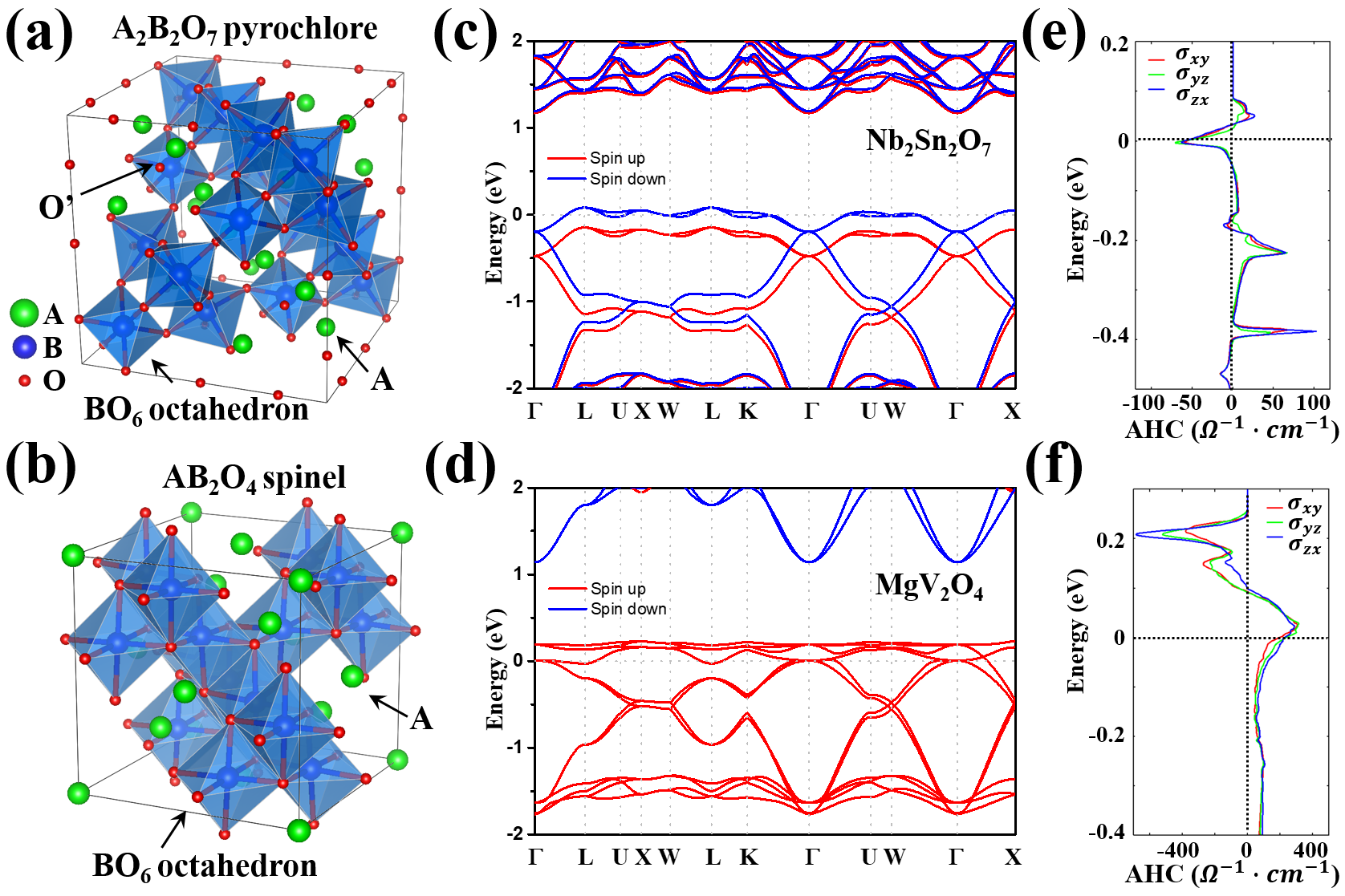} 
\caption{\textbf{Giant AHE in pyrochlore and spinel compounds.} (a) Crystal structure of the pyrochlore compounds with the A cations forming the 3D kagome lattice. (b) Band structure of hole doped Sn$_2$Nb$_2$O$_7$. (c) DFT calculated AHC. (d) - (f) same as (a) - (c) for the  spinel compounds MgV$_2$O$_4$, where cations B form the 3D kagome lattice.}
\label{fig:fig4}
\end{figure}

\paragraph*{Pyrochlore and spinel compounds} It is of great interest to find real materials with high AHC. Therefore, using DFT calculations \cite{Supp}, we seek real materials that can be described by our TB model, which should in principle yield giant AHC. There are two families of compounds, i.e., pyrochlore and spinel compounds~\cite{Hill1979,sickafus1999,Chakoumakos1984,Gardner2010}, that have been extensively studied for decades and could potentially host such lattice model. Though with different chemical formula, both compounds have a group symmetry of $Fd\Bar{3}m$. The crystal structure of the mostly studied $\alpha-$pyrochlore is shown in Fig.~\ref{fig:fig4}(a) with a chemical formula of A$_2$B$_2$O$_7$. It can be viewed as BO$_6$ octahedron clusters that form the pyrochlore sublattice, and A cations forming another set of pyrochlore sublattice. Similarly, the normal spinel compounds with chemical formula of AB$_2$O$_4$ has the structure with BO$_6$ octahedron forming the pyrochlore sublattice and A cations forming a diamond sublattice, as shown in Fig.~\ref{fig:fig4}(b). With proper selection of A and B cations with suitable valence electrons, we will be able to realize the 3D FB and the corresponding giant AHE. 

For the pyrochlore compounds, we choose Nb$_2$Sn$_2$O$_7$ as an example, which has already been demonstrated to host the 3D FBs~\cite{Hase2018,Zhou2019}. However, the spin degenerated 3D FBs are located right below the Fermi level, which need hole doping to partial fill the FBs and trigger the spin splitting according to Stoner's criteria. It is worth mentioning that the FBs have four-fold degeneracy, contributing to even larger density of state than normal FB systems. Therefore, the band structure will experience a large spin splitting with even a small amount of hole doping, as shown in Fig.~\ref{fig:fig4}(c) for the system with one hole. Noticeable band dispersion of those FBs is due to the non-negligible NNN hoppings as we demonstrated earlier on. Then, we turned on the SOC effect and fitted the band structure using Maximally localized Wannier functions, from which we can get the TB Hamiltonian and calculate the AHC, as shown in Fig.~\ref{fig:fig4}(e). As expected, we see a large AHC peak near the FBs right at the Fermi level. However, the peak value is not as large as the TB model, which is due to the large lattice constant of the system as well as the nonvanishing NNN hopping that make the FBs dispersive. We also calculate Nb$_2$Pb$_2$O$_7$, which shows FBs with less dispersion and a corresponding enhancement of AHC \cite{Supp}.

Lastly, we explore the spinel compounds, which are famous for their intriguing magnetic properties \cite{Jiang2020,Hill1979}. One representative experimentally synthesized compound, MgV$_2$O$_4$, is studied, which has also been predicted to host 3D FBs in its ferromagnetic (FM) state~\cite{Azadani2020}. The band structure for the FM MgV$_2$O$_4$ is shown in Fig.~\ref{fig:fig4}(d), which exhibits clear half metallicity with various FBs right above the Fermi level. Interestingly, two sets of the bands remain nearly degenerate that contribute to four FBs near the Fermi level with the bandwidth much narrower than that in pyrochlore compounds. Considering the higher degeneracy of FBs and their narrower bandwidth as well as a smaller lattice constant, we expect a further enhancement of the AHE. The calculated AHC of MgV$_2$O$_4$ is indeed much larger than pyrochlore compounds, as shown in Fig.~\ref{fig:fig4}(f). We also calculate other spinel compounds that have similar band structure, which could all yield giant AHC \cite{Supp}. 

\paragraph*{Discussion and perspectives} It is generally believed that AHE is proportional to the magnetization or strength of exchange splitting~\cite{Nagaosa2010,Thakur2020}. Here, based on a Weyl semimetal toy model, we demonstrate that the microscopic hopping could also be engineered to enhance the AHE by increasing $K$, i.e., reducing the electronic bandwidth of Weyl-related bands. Normally, large SOC is needed to achieve large AHE, as reported for most of the heavy magnetic metals~\cite{Nagaosa2010}. Here, utilizing the Weyl features induced by double-degenerate 3D FBs, it is possible to achieve the giant AHE even in small SOC compounds, such as Nb$_2$Sn$_2$O$_7$. Considering the generality of the pyrochlore model and diversity of pyrochlore and spinel compounds, we expect to greatly expand the number of material candidates with giant AHE. On the other hand, we demonstrated the formation of doubly-degenerate 3D FBs only in pyrochlore lattice, which yields an intriguing giant AHE due to the nearly vanishing bandwidth and superimposition of various Weyl pairs. We believe such phenomenon could also be generalized to other lattice models with 3D degenerate FBs, which deserves further studies.

\textit{Acknowledgement.} This project is supported by SMART, one of seven centers of nCORE, a Semiconductor Research Corporation program, sponsored by National Institute of Standards and Technology (NIST). We acknowledge the MSI in the University of Minnesota for providing the computational resources.

%

\end{document}